\documentclass[apj]{emulateapj_NEI}

\tightenlines

\usepackage{ulem} 
\usepackage{color} 
\usepackage{epsfig,amsmath}
\usepackage{times}

\newcount\listnorom
\newcommand\listromanDE{\global\advance \listnorom by 1
{\lowercase\expandafter{(\romannumeral\listnorom)}\ }}
\newcommand\newlistroman{\listnorom=0}

\newcommand{\RH}{Rankine-Hugoniot}

\newcommand{\fApp}{f_\mathrm{trap}}
\newcommand{\DelT}{\Delta t}
\newcommand{\fSA}{f_\mathrm{SA}}
\newcommand{\FS}{forward shock}
\newcommand{\RS}{reverse shock}
\newcommand{\rgIndex}{{\alpha_\mathrm{rg}}}
\newcommand{\nIndex}{\beta_n}

\newcommand{\RadMax}{R_\mathrm{max}}

\newcommand{\Mshell}{M_\mathrm{shell}}
\newcommand{\rg}{r_g}
\newcommand{\BCSMz}{B_\mathrm{CSM,0}}
\newcommand{\DCSM}{D_\mathrm{CSM}}
\newcommand{\DCSMz}{D_\mathrm{CSM,0}}
\newcommand{\LCSM}{\lambda_\mathrm{CSM}}
\newcommand{\Lz}{\lambda_\mathrm{CSM,0}}
\newcommand{\pmax}{p_\mathrm{max}}
\newcommand{\Qesc}{Q_\mathrm{esc}}

\newcommand{\NCSM}{n_\mathrm{CSM}}
\newcommand{\Nuni}{n_\mathrm{uni}}
\newcommand{\Nshell}{n_\mathrm{shell}}

\newcommand{\Rshell}{R_\mathrm{shell}}
\newcommand{\Rsk}{R_\mathrm{sk}}

\newcommand{\cutoff}{\alpha_\mathrm{cut}}

\newcommand{\jCRmean}{\overline{j^\mathrm{cr}}}

\newcommand{\Mtot}{M_\mathrm{tot}}
\newcommand{\pcc}{cm$^{-3}$}
\newcommand\Msun{\mathrm{M}_{\odot}}

\newcommand{\Bamp}{B_\mathrm{amp}}
\newcommand{\EffDSA}{{\cal E_\mathrm{DSA}}}

\newcommand{\Kep}{K_\mathrm{ep}}

\newcommand\tSNR{t_\mathrm{SNR}}
\newcommand\dSNR{d_\mathrm{SNR}}
\newcommand\EnSN{E_\mathrm{SN}}
\newcommand\Mej{M_\mathrm{ej}}
\newcommand{\RFS}{R_\mathrm{FS}}

\newcommand{\fsk}{f_\mathrm{sk}}

\newcommand{\SA}{semi-analytic}
\newcommand{\NT}{non-thermal}

\newcommand{\CD}{contact discontinuity}
\newcommand{\xx}[1]{\!\times\!10^{#1}}

\newcommand{\Fesc}{F_\mathrm{esc}}
\newcommand{\FescP}{F_\mathrm{esc}(p)}
\newcommand{\DSA}{diffusive shock acceleration}
\newcommand{\CSM}{circumstellar medium}
\newcommand{\MFA}{magnetic field amplification}
\newcommand{\rgz}{r_{g,0}}

\newcommand{\kmps}{km s$^{-1}$}
\newcommand{\NL}{nonlinear}
\newcommand{\gamray}{$\gamma$-ray}
\newcommand{\gamrays}{$\gamma$-rays}
\newcommand{\usk}{V_\mathrm{sk}}

\newcommand{\Lfeb}{L_\mathrm{FEB}}

\newcommand{\SNRJ}{SNR RX J1713.7-3946}

\newcommand{\SC}{self-consistent}
\newcommand{\SCly}{self-consistently}

\newcommand{\muG}{$\mu$G}

\newcommand{\BCSM}{B_\mathrm{CSM}}

\newcommand{\be}{\begin{eqnarray}}
\newcommand{\ee}{\end{eqnarray}}

\newcommand{\DSAinj}{\eta_\mathrm{inj}}

\newcommand{\rel}{relativistic}
\newcommand{\nonrel}{non\-rel\-a\-tiv\-is\-tic}

\newcommand{\ultrarel}{ul\-tra-rel\-a\-tiv\-is\-tic}

\newcommand{\mc}{Monte Carlo}
\newcommand{\MC}{Monte Carlo}
\newcommand{\syn}{synchrotron}
\newcommand{\synch}{synchrotron}
\newcommand{\pion}{pion-decay}
\newcommand{\IC}{inverse-Compton}
\newcommand{\brems}{bremsstrahlung} 

\newcount\Inum
\newcount\IInum
\newcount\IIInum
\newcount\IVnum
\def\I{\global\multiply\IInum by 0 \global\multiply\IIInum by 0
            \global\multiply\IVnum by 0 \global\advance \Inum by 1
            {\the\Inum. }}
\def\II{\global\multiply\IIInum by 0\global\multiply\IVnum by 0
       \global\advance \IInum by 1 {\the\Inum.\the\IInum. }}
\def\III{\global\multiply\IVnum by 0\global\advance \IIInum by 1
            {\the\Inum.\the\IInum.\the\IIInum. }}
\def\IV{\global\advance \IVnum by 1
            {\the\IVnum. }}
\begin{document}

%
\title{Gamma-ray emission of
accelerated particles escaping a supernova remnant in a molecular
cloud}

\author{Donald C. Ellison\altaffilmark{1} and Andrei M. Bykov
\altaffilmark{2}}

\altaffiltext{1}{Physics Department, North Carolina State
University, Box 8202, Raleigh, NC 27695, U.S.A.;
don\_ellison@ncsu.edu}

\altaffiltext{2}{Ioffe Institute for Physics and Technology, 194021
St. Petersburg, Russia; byk@astro.ioffe.ru}

\begin{abstract}
We present a model of gamma-ray emission from core-collapse supernovae
originating from the explosions of massive young stars.  The fast
forward shock of the supernova remnant (SNR)
can accelerate particles by diffusive
shock acceleration (DSA) in a cavern blown by a strong, pre-supernova stellar
wind. As a fundamental part of \NL\ DSA, some fraction of the
accelerated particles escape the shock and interact with a surrounding
massive dense shell producing hard photon emission.
To calculate this emission, we have developed a new \mc\ technique for
propagating the cosmic rays (CRs) produced by the \FS\ of the SNR, into the dense, external material.  This technique is
incorporated in a hydrodynamic model of an evolving SNR which includes
the \NL\ feedback of CRs on the SNR evolution, the production of
escaping CRs along with those that remain trapped within the remnant,
and the broad-band emission of radiation from trapped and escaping CRs.
While our combined CR-hydro-escape model is quite general and applies to
both core collapse and thermonuclear supernovae, the parameters we
choose for our discussion here are more typical of SNRs from very
massive stars whose emission spectra differ somewhat from those produced
by lower mass progenitors directly interacting with a molecular cloud.
\end{abstract}


\section{Introduction}
Many core-collapse supernovae are expected to explode within their
parent molecular clouds.  Because of the influence of the surrounding
material, the manifestation of the supernova remnant (SNR) can differ
substantially depending on the progenitor star type.
For a relatively low progenitor mass below $\sim 12$-14\,$\Msun$, the
stellar wind and photoionizing radiation are not sufficient to
substantially clear out the surrounding cloud and already at a radius
of about 6 pc the remnant can enter a radiative phase with a shock
directly interacting with the molecular cloud 
\citep[e.g.,][]{Chevalier99}. The radiative shock with a typical velocity below $\sim 150$\,\kmps\ can accelerate and compress CRs and produce non-thermal radiation \citep[][]{bceu00,uchiyamaea10}.  Recently, the {\it Large Area Telescope} on board the {\it Fermi Gamma-Ray Space Telescope} detected GeV emission from SNRs IC~443, W44, and 3C\,391 known to be directly interacting with molecular clouds 
\citep[see, e.g.,][]{AbdoEtalIC4432010,AbdoEtalW442010,cs10}.

Higher mass young stars with  masses above  $\sim 16\,\Msun$ (of
B0~V type and earlier) are likely to create low-density bubbles and
HII regions with radii $\sim 10$\,pc surrounded by a massive shell
of matter swept up from the molecular cloud by the wind and the
ionizing radiation of the star over its lifetime.
In this case, a strong supernova shock propagates for a few thousand
years in tenuous circumstellar matter with a velocity well above
$10^3$\,\kmps\ before reaching the dense massive shell where it
decelerates rapidly.

Regardless of the SN type, the blast wave of the SNR is expected to
accelerate ambient material and generate \rel\ electrons and ions,
i.e.,
cosmic rays (CRs), which produce strong non-thermal radiation. A
preponderance of evidence suggests that the particle acceleration
mechanism most likely responsible is \DSA\ (DSA)
\citep[e.g.,][]{BE87,JE91,MD2001}.

We note that despite the common acceleration mechanism, the appearance
of the two classes of SNRs we mention can differ very substantially.
For early progenitor stars, one can expect that a sizeable fraction of the \gamray\ emission is produced by the CR ions that
escape the forward shock and interact with the dense surrounding shell, while for lower mass progenitors, the bulk of the non-thermal radiation is likely to come from trapped CRs.

While the CRs produced by the SNR generate \NT\ emission across the
spectrum from radio to TeV \gamrays, the \gamrays\ are of
particular interest because they may be produced in proton-proton (or heavier ion) collisions of \ultrarel\ particles. In fact, there are three populations
of shock accelerated CRs that are important for producing \gamrays:
relativistic electrons producing \gamrays\ through \IC\ and \NT\
\brems; CR ions that remain trapped within the \FS\ precursor; and CR ions that are accelerated by the \FS\ but escape upstream.
These three populations are produced simultaneously by DSA but they have very different properties and will have very different \gamray\ signatures.\footnote{A fourth particle population that we don't consider here are secondary electron-positron pairs produced by proton-proton interactions \citep[see, for example,][]{GAC2009}. These leptons will produce \IC\ emission and may be important depending on the external mass concentration.}

As has been known for some time
\citep*[e.g.,][]{EJE1981,Eichler84,EE84,BK1988}, a large fraction of the energy in particles accelerated at strong shocks can escape at an
upstream boundary.
In fact, the fraction of all galactic CRs that originate as escaping
particles is likely to be significant and escaping CRs may even
provide
the bulk of CRs at the ``knee'' and above.  The importance of modeling
escaping CRs was discussed before the advent of DSA
\citep[e.g.,][]{ss78} and is attracting considerable attention
currently within the DSA paradigm 
\citep[see, e.g.,][]{PZ2005a,CAB2010,Drury2010}.
Recently, \citet*{rkd09} used a simple iterative scheme to
construct stationary numerical solutions to the coupled kinetic and
hydrodynamic DSA equations. The stationary solutions with efficient
acceleration were found when the escape boundary was placed at the
point where the growth and advection of strongly driven, non-resonant
waves where in balance. For that particular case, they derived the
energy dependence of the distribution function close to the energy
break. As we shall argue below, some additional factors, e.g., stochastic Fermi acceleration on long-wavelength fluctuations,
can affect the spectral shape of the escaping particles.

A number of stationary, \NL\ (NL) models of DSA can provide the
integrated escaping CR energy flux as a fraction of the
parameterized overall acceleration efficiency, but no model is yet
able to determine the spectral shape of escaping CRs taking into
account the \SC\ production of magnetic instabilities produced by
both the trapped CRs in the shock precursor and the escaping
CRs.\footnote{In principle, particle-in-cell (PIC) simulations can
solve this problem exactly. However, it must be noted that PIC
simulations cannot yet solve the full NL shock problem for SNRs.
Most current efforts with PIC simulations have been directed toward
\rel\ shocks \citep[e.g.,][]{Spitkovsky2008b,RiqSpit2010}. Shocks in
SNRs are \nonrel\ and \nonrel\ shocks are harder to simulate than
\rel\ ones.
The NL acceleration of particles, both electrons and protons, that
will
produce radiation spanning radio to \gamrays, requires an extremely
large dynamic range that no PIC simulation can yet achieve (at least not in three-dimensions), and these
simulations will not be able to produce results that
can be compared to broadband continuum emission from SNRs for the
foreseeable future. Approximate methods, such as we describe here, are
needed \citep[see][ for a full discussion]{VBE2008}.}
Such a treatment is not yet feasible, creating an important problem
since the interpretation of the \gamray\ emission from young SNRs depends critically on the
uncertain spectral shape of both the trapped and escaping CRs.
Therefore, a suitable
parameterization of the shape of the escaping CR flux is needed to
allow
comparisons with \gamray\ observations of young SNRs in the hope of
constraining NL DSA models.

It is important to note, of course, that SNRs are not stationary and the dynamics of an expanding remnant, even in the simplest spherical case, adds an additional factor to the issue of CR escape. As the remnant expands, the precursor region 
beyond the forward shock that is filled with CRs expands producing a ``dilution" of CR energy density. This effect has been studied in detail by Berezhko and co-workers \citep[e.g.,][]{BEK1996a,BEK1996b}
\citep[see also][]{Drury2010}. In a real shock, the dilution effect is coupled to escape since the lowering of the CR energy density results in less efficient generation of magnetic turbulence and this will change the escape of CRs. Both the flow of energy out of the shock by escape and the dilution of the CR energy density influence the \RH\ conservation relations in similar ways. Both act as energy sinks and both result in an increase in the shock compression and other \NL\ effects. 
In the stationary, plane-shock  approximation for DSA used here, we ignore  dilution and only include the escape of CRs at an upstream free escape boundary. It has been shown, however, that this plane-shock approximation gives essentially the same results for the shock structure as in a spherical, expanding shock if the specific mode of escape is unimportant \citep[see][]{BE99}. The mode of escape becomes important, however, if the escaping CR flux is used to calculate \gamray\ emission in material external to the shock.
%
%
Since we neglect dilution, our escaping CR fluxes, and the \gamray\ emission we calculate from them, are over estimated.

In this paper, we present a new \MC\ technique for propagating
escaping
CRs and calculating their \gamray\ production via \pion, in the \CSM\
(CSM) surrounding the outer SNR blast wave.
This treatment of escaping CRs is added to our CR-hydro simulation
\citep[e.g.,][ and references therein]{EDB2004,EPSR2010} producing a
coherent model where a number of related elements of the SNR
are treated more or less
\SCly.
The hydrodynamic simulation couples the
efficient production of CRs to the SNR evolution, including the
production of escaping CRs as an intrinsic part of the DSA process.
The escaping CRs are emitted from the \FS\ as the SNR evolves, their
propagation is followed as they diffuse in the CSM, and the \gamray\
emission of these CRs is calculated consistently with that from the
CRs
(protons and electrons) that remain trapped within the remnant.
While a number of important approximations are required, 
including the neglect of precursor CR dilution,
this
model represents a fairly complete and internally \SC\ description of
a SNR interacting with a non-homogeneous CSM.

The importance of freshly made CRs interacting with their local
environment to produce \gamrays\ has been recognized for some time and
an extensive literature exists in this field.
In a generalization of the model of \citet{GA2007} \citep[and previous
work, e.g.,][]{AharA1996}, \citet{GAC2009} calculate the broad-band
emission, from radio to TeV \gamrays, from CRs produced by a SNR
interacting with a nearby molecular cloud. They emphasize that,
depending on the parameters, the \gamray\ emission can exceed other
bands by a large factor, suggesting that some unidentified TeV sources
might be associated with clouds illuminated by nearby SNRs.
\citet{GAC2009} also note the importance of the shape of the \gamray\
spectrum for identifying GeV-TeV sources.

The model used by
\citet{GAC2009} is based on that of \citet{PZ2005a} and includes a
description of the evolution of the SNR and the spectrum of escaping
CRs. In \citet{GAC2009}, the important parameter $\pmax$, the maximum
cutoff momentum for the CR spectrum, is parameterized as $\pmax (t)
\propto t^{-\delta}$, where $t$ is the age of the SNR and $\delta$ is
taken to be $\sim 2.48$ to match the CR data below the knee, as measured at Earth.  
Our parameterization of $\pmax$ differs considerably from this as we discuss below.
An important result of \citet{PZ2005a} \citep[see][ for an earlier derivation]{BK1988} that is
incorporated in the \citet{GAC2009} model and is not modeled here  
is that, when integrated over the whole
Sedov phase, the total CR spectrum is  a power law of the form $\Fesc \propto p^{-4}$.
Recently, \citet{CasanovaEtal_J1713_2010} have applied the
\citet{GAC2009} model to \SNRJ\ taking into account the details of the
ambient gas distribution.

The model presented here is similar to that of \citet*{LKE2008}. In
both
cases, the evolving SNR is modeled with a spherically symmetric
hydrodynamic simulation where the efficient production of CRs via DSA
is
coupled to the remnant dynamics. The main difference is in the
treatment
of the diffusion of escaping CRs in the region beyond the SNR \FS.
The work of \citet{LKE2008} uses a ``boxel'' technique whereby, at
each
time step and each spatial grid in the 3D simulation box, particles
are
exchanged between the adjacent boxels according to the particle
momentum, location, and density gradient. In the model presented here,
we use a \mc\ technique to propagate escaping CRs in the region beyond
the \FS. These two methods of propagation have distinct advantages and
disadvantages, and both differ importantly from more analytic models
based on a direct solution of a diffusion equation. In any case, we
feel
the problem of CRs produced by relatively young SNRs interacting with
dense, local material is important enough to be considered with a
variety of complementary techniques.

Other differences between the boxel model of \citet{LKE2008} and our new
\mc\ model are all based on recent refinements of the CR-hydro model
\citep[see][ and references therein]{EPSR2010} and on how we
parameterize the escaping CR distribution, described below.  While these
refinements are important for modeling specific remnants, they do not
substantially change the results given in \citet{LKE2008}.

\section{Model}
\label{sec:model}
The model we present here consists of two main parts.  The CR-hydro part
is used to calculate the evolution of a SNR and is essentially the same
as that described in \citet{EPSBG2007}, \citet{PES2009},
\citet{EPSR2010} and references therein.
The evolution of the spherically symmetric remnant is coupled to the
efficient production of CRs and the production of thermal and
non-thermal emission is calculated \citep[see][ for recent work modeling
the broad-band emission from \SNRJ]{EPSR2010}.  The \DSA\ is determined
in the CR-hydro model using the \SA\ model of Blasi and co-workers
\citep[e.g.,][]{Blasi2002,AB2005,BGV2005}.  
The injection scheme
for this model has been discussed in detail in a number of previous
papers \citep[see][ for recent extensions of the model]{CAB2010} but we note that we use a slightly different injection method than typically used by Blasi and co-workers. Since the diffusion approximation upon which the \SA\ model is based doesn't apply to thermal particles, a parameter, $\DSAinj$, must
be defined that specifies what fraction of thermal particles obtain a
superthermal energy and are injected into the DSA mechanism. Given this parameter, the \NL\ DSA mechanism determines the fraction of shock ram kinetic energy that goes into superthermal particles, i.e., the acceleration efficiency $\EffDSA$. 
The only difference in our implementation of this injection model and that of Blasi and co-workers is that we specify $\EffDSA$ and then set $\DSAinj$ accordingly. Both schemes are approximations since, in an evolving SNR, both $\DSAinj$ 
and $\EffDSA$ are likely to be functions of age. For simplicity, we hold $\EffDSA$ constant.

The Blasi et al. model that we use also implicitly assumes that the shock is planar and stationary. 
Apart from the neglect of CR dilution,\footnote{We note that, as for other aspects of DSA, CR dilution
will depend importantly on the propagation/acceleration model assumed for the highest
energy CRs. The exact modeling of the highest energy CRs is not yet feasible and we  parameterize all escape effects with our single parameter $\fsk$ defined below.} 
this 
approximation will be reasonably accurate as long as the diffusion length of the highest energy CRs is a small fraction of the shock radius.
The sharp X-ray \syn\ edges 
often seen in SNRs \citep[e.g.,][]{WarrenEtal2005,EriksenEtal2011} implies the presence of amplified magnetic fields which will result in short diffusion lengths. In our models here we assume that the diffusion length of protons with maximum momentum $\pmax$ is 1/10 of the shock radius, a small enough value to validate the planar approximation yet allow $\pmax \sim 10^{4-5} \, m_p c$, consistent with most models of CR production in SNRs.

Accounting for escaping CRs is essential in efficient DSA and escaping
CRs are implicitly included in Blasi's \SA\ description. However, until
now we have not included them in the production of radiation in the
remnant environment in our CR-hydro model. The neglect of radiation
produced by escaping CRs is justified if the SNR is in a uniform CSM
with no external density enhancements. In this case, the emission from
trapped CRs interacting with the shocked material is always much greater
than that produced by escaping CRs in the less dense, unshocked external
medium (see Model B in Fig.~\ref{fig:phot1}).

The second and new part of our model is a calculation of the escaping
CR distribution that emerges from the SNR \FS\ and the propagation and
interaction of these escaping CRs in a dense, spherically symmetric
shell external to the SNR. Depending on the density of the external
material, \gamrays\ produced by the escaping CRs can overwhelm those
produced by trapped CRs, as emphasized by \citet{GAC2009}. We note
that
while here we restrict ourselves to spherical symmetry for the
external
mass distribution, it is straightforward to generalize the \mc\
technique to arbitrary mass distributions.

\subsection{Escaping CR Distribution}
As we make clear in describing our parameterized escaping CR model, both the fraction of energy in escaping CRs and their spectral shape are uncertain. However, 
while controversial for some
years, the idea that some fraction of the most energetic particles
in a shock undergoing DSA must escape, regardless of whether the
shock is stationary or not, is now generally accepted although
certain qualifications are still made \citep[see][]{Drury2010}.
 
We believe that energetic particle escape is a fundamental and unavoidable part 
of DSA that must occur in all supercritical collisionless shocks  regardless of $\pmax$ or time evolution because
(1) observations and modeling of the Earth bow shock 
\citep[e.g.,][]{ScholerEtal1980,MitchellEtal1983,EMP90} support escape, (2) particle escape is an intrinsic part of many particle-in-cell (PIC) simulations 
\citep[e.g.,][]{GBSEB97,GE2000}, and (3) DSA 
requires self-generated turbulence to work over any reasonable dynamic range. Since
CRs must interact with self-generated turbulence to be further accelerated, the highest energy CRs far upstream in the shock precursor will always lack sufficient turbulence to remain nearly isotropic and some fraction will escape.  
These escaping CRs will generate turbulence for the next generation of
CRs, creating a bootstrap effect.
As mentioned above, the dilution of CR energy density that  occurs in spherical, expanding shocks will be coupled to CR escape through the magnetic turbulence generation.

Given the assumptions and approximations of the model, the
 \SA\ description of \citet{BGV2005} determines the energy in
escaping CRs, $\Qesc$, but does not determine the shape of the
distribution. While other work does determine the shape
\citep[e.g.,][]{VEB2006,ZP2008,CAB2010}, the shape that results
in these models depends importantly on arbitrary parameters and the
assumptions made for the diffusion of the highest energy escaping CRs.

Since the shapes of the trapped and escaping CR distributions, at the
highest accelerated energies, are critical for modeling both X-ray
\syn\ emission and GeV-TeV \gamray\ emission, we feel it is important
to
have a flexible, i.e., parameterized, model that can be compared to
observations to provide information on the uncertain plasma processes
until an adequate theory of self-generated turbulence in the presence
of
escaping particles is developed \citep*[see][ for recent work on
  long-wavelength instabilities that may influence the maximum
momentum CRs can obtain in a given shock]{BOE2011}.

As an example of the complexities that may exist, the amplified
long-wavelength fluctuations discussed in \citet*{BOE2011} may
result in particle acceleration by the resonant second-order Fermi
mechanism. The stochastic acceleration rate, $\tau_{\rm ac}^{-1}$,
for particles with spatial diffusion coefficient $\kappa(p)$ in the
shock precursor is $\tau_{\rm ac}^{-1}\propto v^2_{\rm
ph}/\kappa(p)$, where $v_{\rm ph}$ is the phase velocity.  While
this rate may be below the first-order acceleration rate, it may
still be high enough to influence the spectra shape at the highest
particle energies achieved by first-order DSA. The spectral
index of particles accelerated by the second-order Fermi mechanism
depends on the parameter $\tau_{\rm ac}/T_{\rm esc}$, where $T_{\rm
esc}$ is the escape time \citep[e.g.,][]{pb08}.  In the case of resonant stochastic particle acceleration by long-wavelength fluctuations,
$\tau_{\rm ac}/T_{\rm
  esc} \propto M_{\rm a}^2/ [k_1\,r_g(p_{\rm max})]$,
where the
characteristic wave number of
  the CR instability \citep[c.f.,][]{Bell2004,BOE2011} is
\begin{equation}\label{k1}
k_1 = \frac{4\pi}{c} \frac{\jCRmean}{B} \ .
\end{equation}
Here, $\jCRmean$ is the mean CR current,
$r_g(p_{\rm max})$ is the CR gyroradius at $\pmax$
in the magnetic field $B$, and $M_{\rm a}$ is the
forward shock Alfv\'enic Mach number.

Therefore, for a large enough precursor CR current, $\jCRmean$,
as expected for efficient DSA, the parameter $\tau_{\rm
  ac}/T_{\rm esc}$ may  influence the shape of the CR distribution in the spectral break region. For instance, if a shock of velocity $\usk$ produces a power-law
spectrum of accelerated particles up to some maximum momentum and
transfers a fraction $\eta$ of the shock ram pressure to CRs, then
$\tau_{\rm ac}/T_{\rm esc} \propto \eta^{-1}\, (c/\usk)\,$ with a weak dependence on the particle
  momentum.
The smaller $\tau_{\rm ac}/T_{\rm esc}$, the larger is the second-order Fermi effect and preliminary work (A. Bykov, in preparation) suggests that  $\tau_{\rm ac}/T_{\rm esc} \lesssim 100$ is needed to see a significant modification of the spectral shape. While more exact estimates are difficult, we might expect $\eta \gtrsim 0.5$ and $\usk \gtrsim 5000$\,\kmps\ to produce a noticeable effect.


When the shock accelerated particles approach $\pmax$, they begin
leaving the upstream region of the shock and the approximate
power-law distribution of particles that remain in the shock turns
over in a fashion that will depend on the diffusion coefficient of
the highest energy particles.
%
Whatever the plasma processes are for escaping particles, the shape
of the escaping distribution, $\FescP$, is determined by how CRs
leave the shock and is, therefore, coupled to the trapped
distribution. Since no current model of self-generated turbulence
adequately describes the diffusion of escaping CRs,  the diffusion
coefficient is generally assumed to be Bohm-like right up to
$\pmax$ and independent of position relative to the shock.

\begin{figure}
\epsscale{0.9}
\plotone{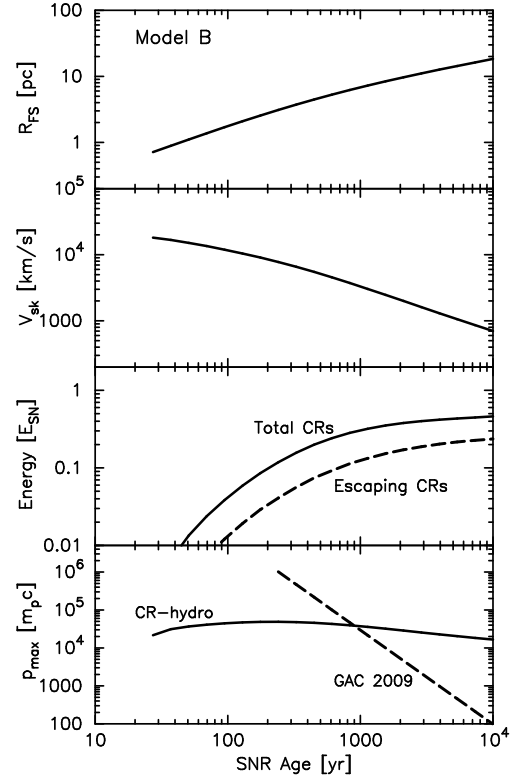} 
\caption{The curves in the top two panels show the forward shock
radius, $\RFS$, and speed, $\usk$, as a function of remnant age. 
Both $\RFS$ and $\usk$ are determined directly from the hydro code and include the energy loss from escaping CRs and all adiabatic effects. The gradual change in slope indicates a broad transition between 
an ejecta-dominated early phase and the Sedov phase at later times. The third panel shows the fraction of SN explosion energy in all CRs along with the fraction going into escaping CRs. In the bottom panel, $\pmax$ from the
CR-hydro simulation (solid curve) is compared to that  used in
\citet{GAC2009} (dashed curve). While the curves in this figure
extend to $10^4$\,yr, the simulations discussed in the remainder
of this paper stop at $\tSNR=1000$\,yr, well before the SNR is fully
in
the Sedov phase. We note that the near identity of the CR-hydro and
\citet{GAC2009} value of $\pmax$ at $1000$\,yr is essentially a
coincidence. The parameters used for these results are listed as
Model~B in Table~\ref{tab:tableA} but the quantities displayed here apply to all of our examples.
\label{fig:pmax}}
\end{figure}

Here, we parameterize the escaping CR phase-space distribution,
$\FescP$, as a modified parabola centered at $\pmax$, where $\pmax$ is
the maximum momentum CRs would obtain if the acceleration cut off
sharply when the upstream diffusion length,
$\kappa(\pmax)/\usk = \Lfeb$,
where $\usk$ is the speed of the FS and $\Lfeb$ is a free escape
boundary.
In our SNR model, the maximum momentum is determined primarily by an
arbitrary parameter, $\fsk$, which is the fraction of the shock radius
equal to the diffusion length of protons with momentum $\pmax$, i.e.,
$\Lfeb(t) = \fsk \Rsk(t)$, where $\Rsk(t)$ is the radius of the FS at
time, $t$. For all of the examples shown here,
we set $\fsk=0.1$; a factor small enough to be consistent with the planar shock approximation in the Blasi et al. DSA 
calculation.\footnote{We note that at early times, setting the acceleration time equal
to the age of the remnant may give a lower $\pmax$ in which case this
value is used \citep[see, for example,][]{BEK1996a,EDB2004}.}

\begin{figure}
\epsscale{0.9}
\plotone{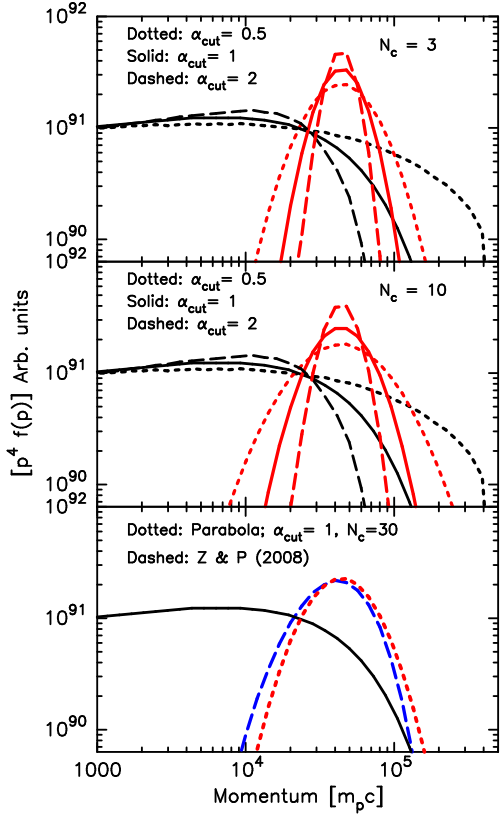} 
\caption{The top two panels show trapped (black curves) and escaping
CR
  (red curves) proton distributions for simulations where $\cutoff$
and
  $N_c$ have been varied. The distributions are summed at the end of
the
  simulation at $\tSNR=1000$\,yr and the escaping CR distributions are
  those leaving the FS before any propagation occurs.  In the bottom
  panel we compare our parabola fit with $\cutoff=1$ and $N_c=30$ (red
  dotted curve) to the result from \citet{ZP2008} (blue dashed
  curve). Both curves have been normalized to the total energy in
  escaping CRs.
\label{fig:shape}}
\end{figure}

Our scheme for determining $\pmax$ gives a very different result from
the parameterization used in \citet{GAC2009}, as indicated in the
bottom
panel of Fig.~\ref{fig:pmax}. \citet{GAC2009} argue that \MFA\ (MFA)
may
contribute to a strong decrease in $\pmax$ as a function of time
since it might be expected that MFA is
strongest at early times, yielding a large magnetic field and a
higher $\pmax$. As the remnant ages,
MFA might decrease, producing a stronger time dependence than the
standard SNR evolution would suggest. Magnetic field amplification is
not include in the examples we show here. We caution, however, that
\NL\ feedback may reduce the full effects of MFA
\citep[e.g.,][]{VBE2008,CBAV2008,CBAV2009} and we feel it is unlikely
that
a
time dependence as strong as assumed by \citet{GAC2009} will be
obtained. In any case, our main purpose here is to introduce a new
propagation tool for escaping CRs and not be overly concerned with
details that are still subject to active research.

An approximate expression for the CRs that remain trapped within the
SNR is \citep[e.g.,][]{EDB2004}
%
\begin{equation} \label{eq:fp}
\fApp(p) \sim  \fSA(p)
\exp{\left [ -\left ( \frac{p}{\pmax}\right )^{\cutoff} \right ]}
\ ,
\end{equation}
where $\fSA \sim (p/\pmax)^{-4}$ is the quasi-power law DSA
distribution
obtained by the standard \SA\ model, $\cutoff$ is an arbitrary
parameter
that determines the turnover around $\pmax$, and $\pmax$, as
mentioned,
is determined by the SNR dynamics and $\fsk$.
%
%
The distribution of escaping CRs, $\Fesc(p)$, is parameterized by
assuming that it is a parabola in $\log{(p^4 \Fesc})$ space, that is,
\begin{eqnarray}\label{eq:Fesc}
& &
\log{\left [ (p')^4 \FescP \right ] } =
\nonumber \\
& &
-a\left [ \log{(p')} - \log{(1)}  \right ]^2 + b
\ ,
\end{eqnarray}
where $p'=p/\pmax$. Initially, we determine $b$ such that
\begin{equation}
\fApp(\pmax) = \Fesc(\pmax)
\ ,
\end{equation}
which yields $b=\log(e^{-1}) = -0.434$.

\begin{figure}
\epsscale{1.0}
\plotone{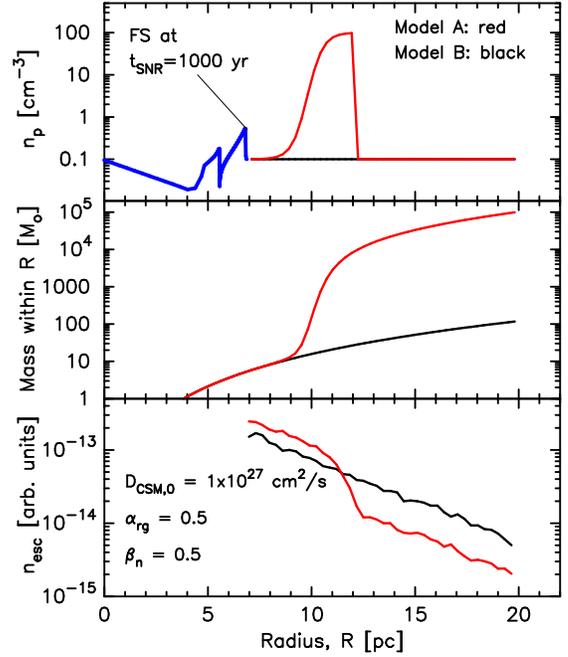} 
\caption{The black and red curves in the top panel show two
spherically
symmetric CSM density profiles. The blue curve in the top panel shows
the density of the SNR at $\tSNR=1000$\,yr. The two curves in the
middle
panel show the total mass within a particular radius for the CSM
profiles at the start of the simulation. The dense shell has a mass of
$\Mtot = 10^4\,\Msun$.  In both cases, CRs escaping from the FS of the
SNR
propagate in the CSM profiles and, at $\tSNR=1000$\,yr, have
the color-coordinated density profiles shown in the bottom panel.
The parameters for the CSM propagation (e.g., Eq.~\ref{eq:mfp}) are
shown in the bottom panel. The density profiles shown in the bottom panel are for a particular momentum near the peak of the escaping CR
distribution. The irregular variations in the escaping CR densities
are a result of the stochastic nature of the \MC\ propagation.
\label{fig:Ext1}}
\end{figure}

The width of the parabola, $a$, is matched to $\fApp(p)$ as follows.
We determine the momentum $p_c > \pmax$ where the trapped CR
distribution drops by some factor, $1/N_c$, below its value at $\pmax$,
i.e.,
\begin{equation}\label{eq:match}
\frac{\fApp(p_c)}{\fApp(\pmax)} = 1/N_c
\ .
\end{equation}
Specifying $N_c$ uniquely determines $p_c$. We then obtain $a$ by
setting:
\begin{equation}
\frac{\Fesc(p_c)}{\Fesc(\pmax)} = N_c
\ ,
\end{equation}
that is,
\begin{equation}
a(\cutoff) = -
\log{\left [ (p'_c)^4 N_c\right ]} / \left [ (\log{p'_c}) \right ]^2
\ ,
\end{equation}
where $p'_c = p_c/\pmax$ and we have written $a(\cutoff)$ to emphasize
that the width of the escaping distribution depends on the cutoff
parameter in the trapped CR distribution. The final normalization for
$\Fesc$ is set by the total energy in the escaping distribution,
$\Qesc$, which is an output of the \SA\ DSA model.

In the top two panels of Fig.~\ref{fig:shape} we show examples where
$\cutoff$ is varied between 0.5 and 2 with $N_c=3$ and $10$.  All
other parameters of the CR-hydro model are the same for these
examples. In all panels, the black curves are the CRs that remain
trapped in the SNR, $f(p)$, and these distributions, unlike the
escaping
CRs, have undergone adiabatic losses during the $\tSNR=1000$\,yr age
of
the remnant.\footnote{We note the distinction that the trapped
distributions, $f(p)$, in these plots determine $\fSA$ exactly from
the CR-hydro model and the \SA\ DSA calculation, as opposed to the
approximate expression for $\fApp$ used in Equation~(\ref{eq:fp})
to fit the modified parabola.}
The parameters, $\cutoff$ and $N_c$ allow a fairly wide range of
shapes in the critical region around $\pmax$, although it is
important to note that, in our model, for all reasonable values of
$\cutoff$ and $N_c$, the escaping CR distribution is expected to be
narrow compared to the trapped CRs. This differs from the work of
\citet{GAC2009}, as mentioned above, and of \citet{OhiraEtal2010}
who assume a power-law form for the escaping CR distribution.

In the bottom panel of Fig.~\ref{fig:shape} we compare our
parameterization (red dotted curve) using $\cutoff=1$ and $N_c=30$ to
the form presented in \citet{ZP2008} (blue dashed curve).  Other than
a
slight offset of the peak, this choice of $\cutoff$ and $N_c$ matches
the \citet{ZP2008} result quite well. We could have obtained an
equally
good match with a different combination of $\cutoff$ and $N_c$.  The
quality of this match with $\cutoff=1$ leads us to fix $N_c=30$ and
leave $\cutoff$ as a single free parameter for the coupled shapes of
the
cutoff in the trapped CRs and the escaping distribution.


\subsection{Monte Carlo Model of Cosmic Ray Propagation}
Given the form for the escaping distribution, we propagate the
escaping CRs using a \MC\ technique.\footnote{Many of the elements of
our \MC\ propagation model are similar to that used to model \NL\
DSA and are described in detail in \citet{JE91} and
\citet*{EBJ96} and references therein.}
As the CR-hydro simulation evolves,
$\Fesc(p)$ is calculated for spherical shells at time-steps, $\DelT$,
as
the \FS\ overtakes fresh circumstellar material. As the outer-most
shell
is formed, escaping CRs leave the shell and diffuse into the CSM with
a
momentum and density dependent mean free path given by
\begin{equation} \label{eq:mfp}
\LCSM = \Lz (\rg/\rgz)^\rgIndex (\NCSM/n_0)^{-\nIndex}
\ .
\end{equation}
Here, $\rg=pc/(eB)$ is the gyroradius, $\NCSM$ is the CSM proton
number
density, and $\rgIndex$ and $\nIndex$ are parameters. For scaling, we
use $n_0 = 1$\,\pcc, $\rgz = 10 \mathrm{GeV}/(e \BCSMz)$, and
$\BCSMz =
3$\,\muG. The normalization of the CSM diffusion coefficient, $\DCSMz=
\Lz \, c/3$, can be estimated from CR propagation studies \citep[see,
for example,][]{PMJSZ2006,GAC2009}. For example, with
$\DCSMz=10^{27}\,$cm$^2$ s$^{-1}$, $\NCSM=0.01$\,\pcc,
$\rgIndex=0.5$, and
$\nIndex=1$, $\LCSM \sim 1$\,pc at 1 GeV, consistent with the fits of
\citet{PMJSZ2006}.

We note that the CSM diffusion resulting from Eq.~(\ref{eq:mfp})
is very different from the diffusion we  assume to occur within the
SNR. For the acceleration process at the FS, we assume Bohm diffusion
with $\lambda \sim \rg$. Once the trapped CRs have been accelerated in
the outer shell, these CRs are assumed to remain in the shell as it
convects and evolves within the remnant. In all cases, the CSM
scattering
is much weaker than within the SNR and the escaping CRs quickly fill
the
CSM out to the end of the simulation box.

The process continues until $\tSNR$ is reached during which time some
number of CR filled shells  have been formed within the SNR. The
escaping
CRs fill the CSM region with a distribution that depends on
Eq.~(\ref{eq:mfp}) and the properties assigned to the CSM.

\subsection{Circumstellar Medium Properties}
\label{sec:CSM}
We model the spherically symmetric CSM with a dense shell sitting on a
low-density, uniform background of density $\Nuni$. The shell has a
maximum density $\Nshell$ and an inner radius $\Rshell$ which, for the
examples in this paper, is greater than the outer radius of the SNR at
$\tSNR$, that is, the blast wave of the SNR has not yet reached the dense shell at the end of the simulation.  An additional parameter is the total mass in the shell,
$\Mshell$.

The red curves in the top two panels of Fig.~\ref{fig:Ext1} show the
density and mass distribution for a CSM with $\Nuni=0.1$\,\pcc,
$\Nshell=100$\,\pcc, $\Rshell \sim 10$\,pc, and
$\Mshell=10^4\,\Msun$. Note that the dense shell smoothly rises from
$\Nuni=0.1$\,\pcc\ and the rise is centered on $\Rshell$.
The black curves show the CSM with no shell.
The total extent of the simulation box for these examples
is $\sim 20$\,pc. Also shown in the top panel (blue curve) is the
density profile of the SNR at the end of the simulation, i.e.,
at $\tSNR=1000$\,yr.  The FS, \CD, and RS can be easily discerned from
the
figure.

The addition of the escaping CR distribution requires additional
parameters and Table~\ref{tab:tableA} gives the parameters for the CSM
diffusion and the parameters for the external medium. As mentioned
above, we restrict ourselves to a spherically symmetric CSM in this
first presentation of the \MC\ propagation model. 

\begin{figure}
\epsscale{0.9}
\plotone{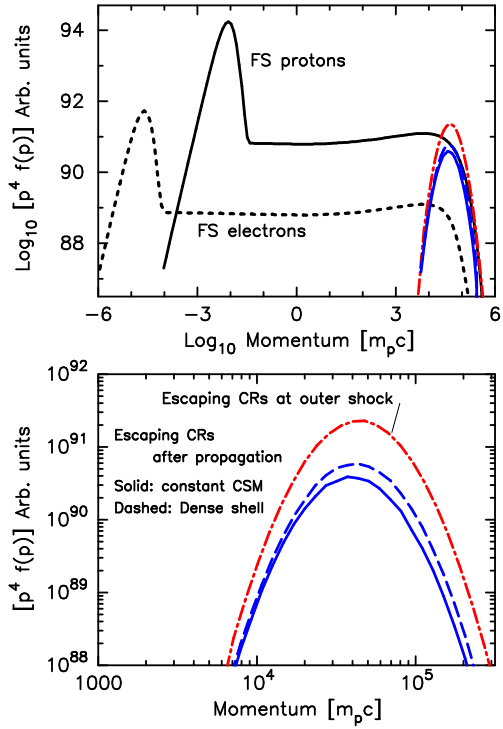} 
\caption{The top panel shows the total CR spectra within the \FS\ at
  $\tSNR=1000$\,yr (black solid curve: protons; black dotted curve:
  electrons), along with the escaping CRs. The red curves in both
panels
  show the escaping CR distribution at the FS, while the blue solid
and
  dashed curves show the escaping CRs after diffusing in the CSM
  profiles shown in Fig.~\ref{fig:Ext1}. The
  solid blue curves (Model B) are for the constant CSM and the dashed
blue
curves (Model A)
  are for the dense shell. The two models in this plot have
  $\rgIndex=\nIndex=0.5$ and $\DCSMz=1\xx{27}$\,cm$^2$/s.
\label{fig:fp1}}
\end{figure}

\begin{figure}
\epsscale{1.0}
\plotone{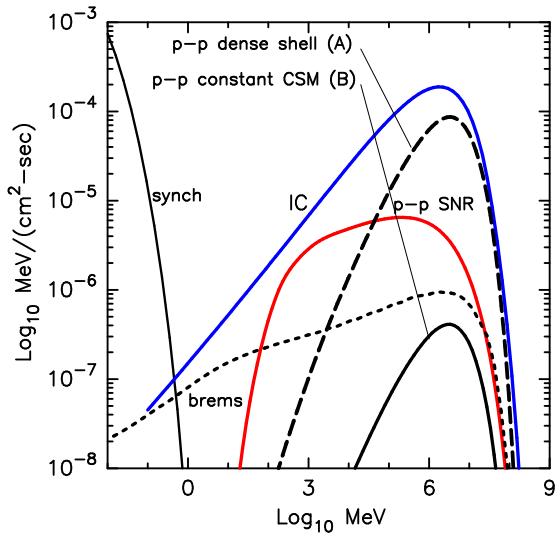} 
\caption{Photon spectra at Earth for the examples shown in
  Fig.~\ref{fig:fp1}. As expected, the case with a dense external
shell (Model A) shines
  much more brightly in \gamrays\ than the case with a low density,
  uniform external CSM (Model B). The individual components for
\synch, \IC,
  \brems, and the \pion\ emission from CRs that remain trapped in the
  SNR are indicated.
\label{fig:phot1}}
\end{figure}

\begin{figure}
\epsscale{0.9}
\plotone{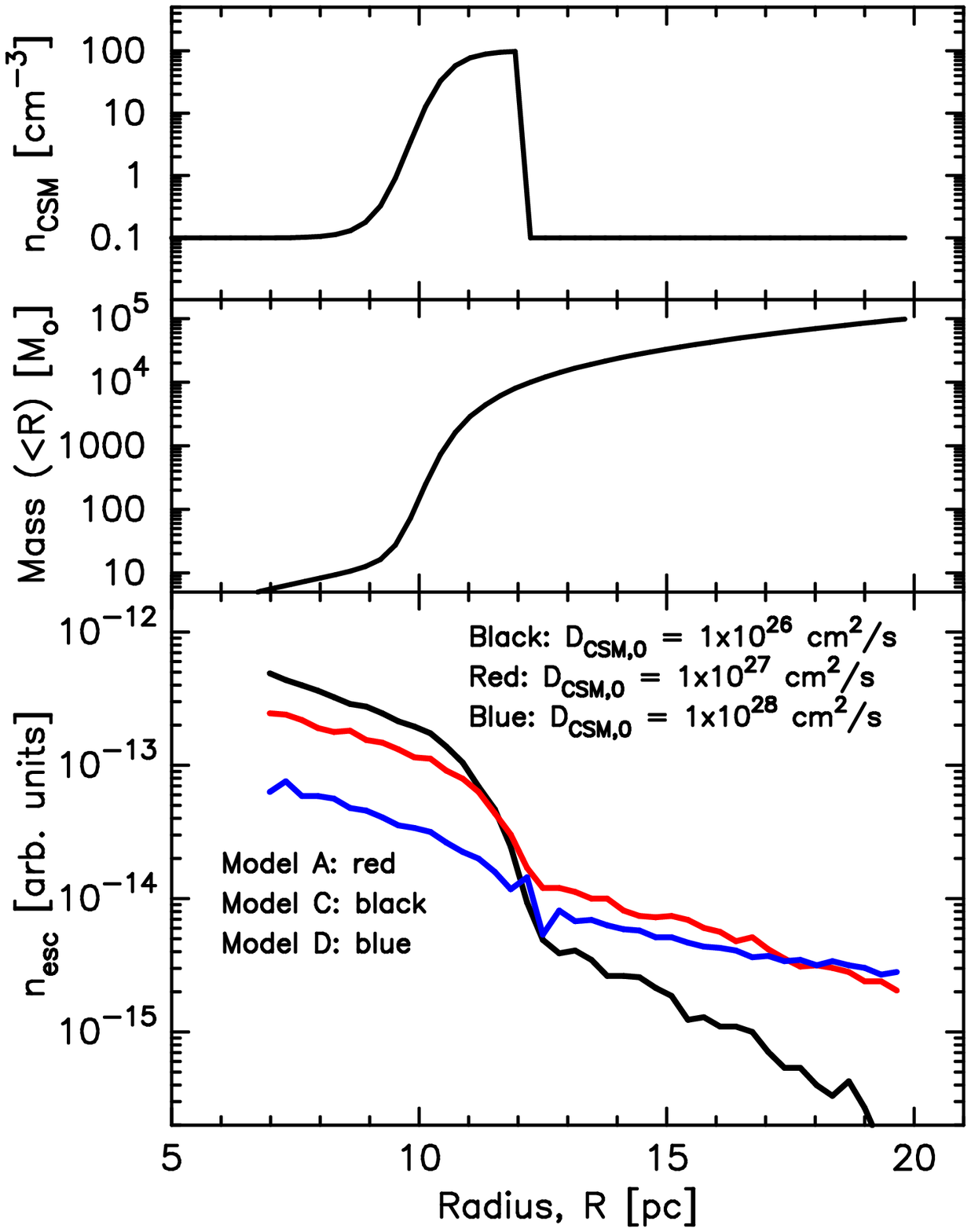} 
\caption{Escaping CR diffusion in a dense external shell as a function
  of the normalization of $\DCSM$. In all cases,
$\rgIndex=\nIndex=0.5$.
\label{fig:ExtVary}}
\end{figure}

\section{Results}
\label{sec:results}

We use the following environmental parameters for all of our examples:
\newlistroman
\listromanDE the SN explosion energy, $\EnSN=10^{51}$\,erg;
\listromanDE the ejecta mass, $\Mej=1.4\,\Msun$;
\listromanDE the distance to the SNR, $\dSNR=1$\,kpc, and;
\listromanDE the ambient magnetic field throughout the CSM,
$\BCSM=3$\,\muG.

For the \DSA\ of trapped and escaping CRs,  we fix the following:
\newlistroman\listromanDE
the fraction of FS radius used to determine $\pmax$, $\fsk=0.1$;
%
%
\listromanDE the magnetic field amplification factor, $\Bamp=1$, i.e.,
no MFA is used;
\listromanDE the matching factor defined in equation~(\ref{eq:match}),
$N_c=30$, and;
\listromanDE the DSA efficiency, $\EffDSA=50$\%.

The parameters for DSA and the CSM propagation that are varied for our
examples are given in Table~\ref{tab:tableA}. Again we note that we
are not attempting a detailed fit to any particular remnant and that our model
is not restricted to the particular values for parameters we use here.
Any of the environmental or DSA parameters can be modified
to match a specific object.

In Fig.~\ref{fig:Ext1} we show results for Models A (with a dense
external shell) and B (no external shell), as listed in
Table~\ref{tab:tableA}.
The top two panels were
discussed in Section~\ref{sec:CSM}.
In the bottom panel of Fig.~\ref{fig:Ext1} we show the escaping CR
densities at $\tSNR=1000$\,yr. The escaping densities shown are for a single momentum near the peak of $\Fesc$ and the
parameters assumed for the CSM propagation are noted in the figure.
The escaping CRs are emitted from the SNR as it evolves so the escaping CRs that were produced earliest have been diffusing for approximately $\tSNR=1000$\,yr and many have left the simulation box.

For the case where the CSM is uniform (black curves), the escaping CRs
diffuse outward and uniformly fill the region beyond the SNR \FS\ with
a density that deceases uniformly with radius as expected.
With the dense shell, the escaping CR density drops rapidly as the CRs
enter the shell.  With this $\Mshell=10^4\,\Msun$,
the shell is about 2\,pc thick. Cosmic rays that propagate beyond
$\RadMax \sim 20$\,pc are removed from the simulation.

In the top panel of Fig.~\ref{fig:fp1} we show the CR distributions
for
both the CRs that remain trapped in the SNR (black solid and dotted
curves) and escaping CRs.\footnote{In all of the examples in this
paper
we only calculate CRs accelerated at the \FS\ and ignore those
accelerated by the \RS.} In all cases, the
integrated distributions are determined at $\tSNR=1000$\,yr.
The red curve in the top panel is the summed escaping
distribution as the CRs leave the FS, i.e., before they propagate into
the external CSM. The solid and dashed blue curves are the escaping
CRs,
at $\tSNR=1000$\,yr, after propagation and we remind the reader that our escaping CR fluxes are over estimates since we don't model dilution which, in fact, occurs simultaneously with escape. The bottom panel shows just
the
escaping CRs with an expanded scale. Note that throughout this paper
we
include only escaping protons and ignore escaping heavier ions and
escaping electrons. Trapped
electrons are considered for \IC\ emission.
The distributions for the escaping CRs after propagation are lower
than
the distribution as CRs leave the FS for two reasons. The first is
that
some CRs escape from the simulation box at $\RadMax$. The second is
that
some escaping CRs
diffuse back into the SNR and these CRs are ignored and not included
in
the blue distributions in Fig.~\ref{fig:fp1}.
The CRs that remain trapped in the shock are summed from the \CD\ to
the
\FS.

In Fig.~\ref{fig:phot1} we show the various photon components for the
models with $\rgIndex=\nIndex=0.5$ and $\DCSMz=1\xx{27}$\,cm$^2$/s.
The
results for the two models are identical except for the \pion\
emission
from the escaping CRs. As expected, escaping CRs interacting with the
dense external shell produce substantially more emission than those
interacting with the uniform CSM. In both cases, however, the emission
from escaping CRs is much more strongly peaked than the \pion\
emission
from the trapped CR protons. For the trapped CRs, the relative
intensity
of the \pion\ emission and the \IC\ emission depends on the various
parameters chosen, most particularly $\Nuni$ and $\Kep$, the electron
to
proton ratio at \rel\ energies. The fact that our values,
$\Nuni=0.1$\,\pcc\ and $\Kep=0.01$,
result in \IC\ dominating the GeV-TeV emission is not necessarily an
indication that we believe this will always be the case. The issue is
more complicated as indicated in a number of recent papers \citep[see,
  for example,][]{KW2008,MAB2009,ZirA2010,EPSR2010}.
Regardless of other parameters, the {\it relative importance} of the
\gamray\ emission from escaping CRs and trapped CRs depends mainly on
the
external density enhancement.

In Fig.~\ref{fig:ExtVary} we show the effect of the normalization of
the
CSM diffusion coefficient as escaping CRs diffuse into the dense
shell. Three effects are noticeable. The first is that the escaping CR
density drops more rapidly with stronger scattering (i.e., smaller
$\DCSMz$) as CRs enter the dense shell. The second is that the
escaping CR density remains larger in the region between the FS and
the
dense shell when scattering is strong even though the flux of escaping
CRs that leave the FS is the same in all three cases.
The third effect is that, beyond the dense shell, the escaping CR
density
falls off faster with stronger scattering.
The CR density remains large
within the shell (i.e., at radii $\lesssim 10$\,pc) for strong
scattering
because the dense shell acts as a valve that slows the
flow of CRs out of the system. Beyond the dense shell, weak scattering
results in a more uniform density distribution than strong scattering
since CRs rapidly fill the available volume when the scattering is
weak.

\begin{figure}
\epsscale{1.0}
\plotone{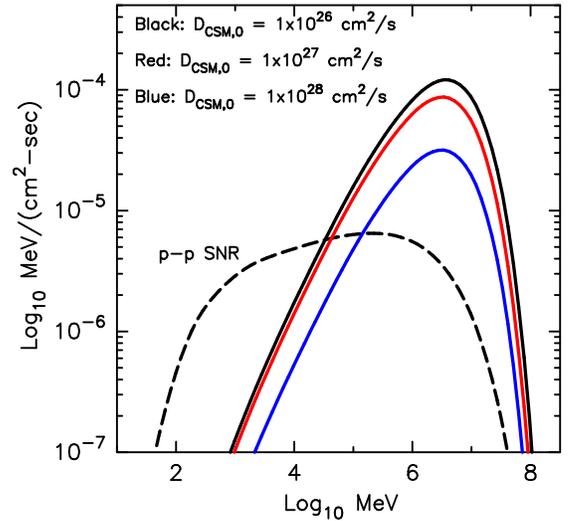} 
\caption{Gamma-ray emission for the three cases shown in
  Fig.~\ref{fig:ExtVary}. Since only the interaction of escaping CRs
  with the external CSM is varied, the \pion\ emission from the
trapped
  CRs within the SNR is the same for the three cases. Referring to
Table~\ref{tab:tableA}, the black solid curve is Model C, the red
solid curve is Model A, and the blue solid curve is Model D.
\label{fig:PhotonVary}}
\end{figure}

In Fig.~\ref{fig:PhotonVary} we compare the \pion\ emission for the
three examples given in Fig.~\ref{fig:ExtVary}, all with the same
ambient density distribution. The CRs trapped in the SNR are the same
for these cases so the \pion\ emission from the trapped CRs (dashed
curve) is the same in the three models.  Also identical for the three
cases is the escaping CR flux as it emerges from the FS.  The sole
difference is the scattering strength, $\DCSMz$, in the CSM and this
produces a fairly strong effect on the \pion\ emission from the
escaping
CRs.
While the emission from escaping CRs shown in Figs.~\ref{fig:phot1}
and
\ref{fig:PhotonVary} is summed over the entire region from the outer
radius of the SNR at $\tSNR$ to $\RadMax \sim 20$\,pc, when a dense
shell is present, most of the emission originates in the shell, as
expected.

\begin{figure}
\epsscale{0.9}
\plotone{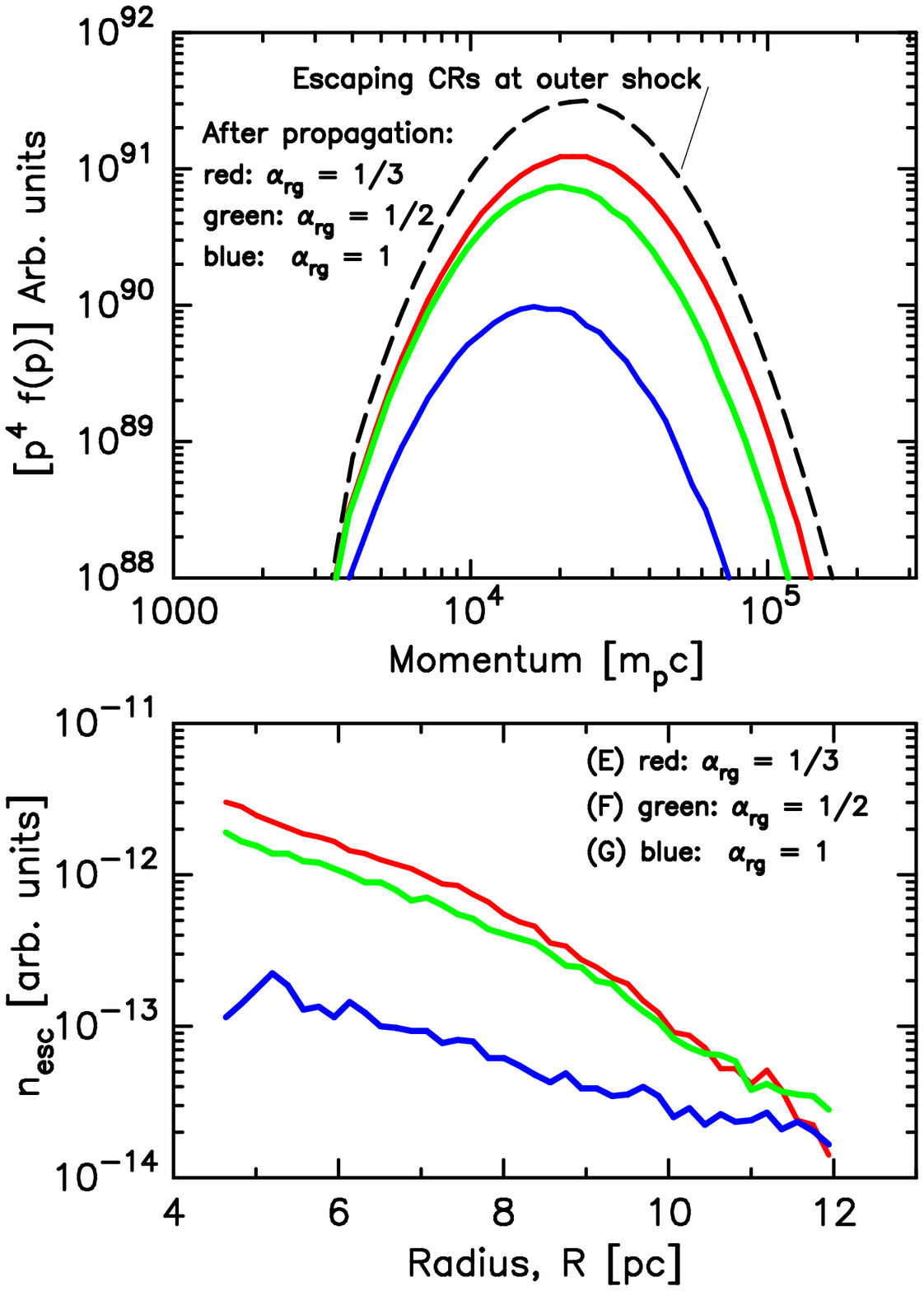} 
\caption{Escaping CR distributions, $\Fesc$, (top panel), and density
profiles (bottom panel) for Models E, F, and G, as listed
in Table~\ref{tab:tableA}.  The index $\rgIndex$ is varied as shown
and $\nIndex=0.5$ in all cases. For these examples, and those shown in
Fig.~\ref{fig:vary_density}, $\Nuni=1$\,\pcc, $\Nshell=10$\,\pcc,
$\Rshell=7$\,pc, and the
densities in the bottom panel are
for a particular momentum near the peak of the escaping CR
distribution.
The simulation box extends to 12\,pc.
\label{fig:vary_gyro}}
\end{figure}

In Fig.~\ref{fig:vary_gyro} we compare escaping CR distributions
for different power-law dependences of the gyroradius, i.e.,
$\rgIndex=1/3$ (Model E), $\rgIndex=1/2$ (Model F), and
 $\rgIndex=1$ (Model G). For variety, the models in Fig.~\ref{fig:vary_gyro},
along with those in Fig.~\ref{fig:vary_density} below,
use a different set of CSM parameters than the models discussed thus
far, as shown in Table~\ref{tab:tableA}.
For the three examples shown, the CSM parameters are identical and the
mean free paths differ only in the value of $\rgIndex$; the
normalization of the diffusion coefficient $\DCSMz=10\xx{27}$\,cm$^2$
s$^{-1}$ and $\nIndex=0.5$ are the same for the three values of
$\rgIndex$.
As the dependence on $\rgIndex$ increases, the high momentum CRs are
able to stream through the CSM quickly and the number that remain
within the simulation region at $\tSNR=1000$\,yr drops.
The strong $\rgIndex$ dependence also results
in a flatter radial density distribution, as indicated by the blue curve in the bottom panel of Fig.~\ref{fig:vary_gyro}. The reason for this is that the momentum near the peak in the escaping distribution that is used to calculate the density profiles is well above 10\,GeV so the examples with larger $\rgIndex$ have longer mean free paths.

In Fig.~\ref{fig:vary_density} we show a similar plot where we now
keep $\rgIndex=0.5$ and vary the power-law
index for the density dependence of the diffusion coefficient,
$\nIndex$. When
$\nIndex=0$ and there is no density
dependence for the diffusion coefficient, the presence of the external
dense shell produces no effect and the red curve in the bottom panel
of  Fig.~\ref{fig:vary_density} falls off uniformly with radius. For
stronger density dependences (green and blue curves), the density of
escaping CRs drops as they enter the dense external shell which has a
radius $\Rshell \simeq 7$\,pc for these models and those shown in Fig.
~\ref{fig:vary_gyro}.

\begin{figure}
\epsscale{0.9}
\plotone{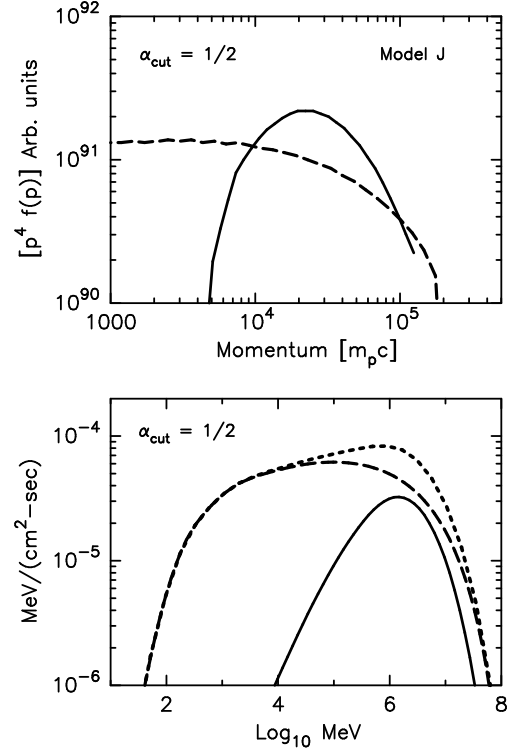} 
\caption{The top panel shows particle spectra for trapped CRs (dashed curve) and escaping CRs (solid curve). The bottom panel shows the corresponding \pion\ emission for these distributions along with the sum (dotted curve).
\label{fig:fpCutHalf}}
\end{figure}

\begin{figure}
\epsscale{0.9}
\plotone{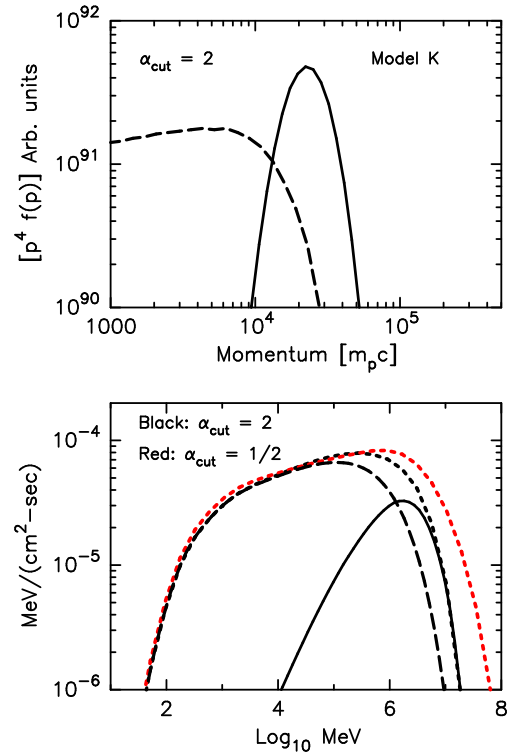} 
\caption{Same as in Fig.~\ref{fig:fpCutHalf} with $\cutoff=2$. In the bottom panel we compare the summed emission for the two cases $\cutoff=1/2$ (red dotted curve) and $\cutoff=2$ (black dotted curve).
\label{fig:fpCutTwo}}
\end{figure}

In Figs.~\ref{fig:fpCutHalf} and \ref{fig:fpCutTwo} we compare \pion\ emission for $\cutoff=1/2$ (Model J) and $\cutoff=2$ (Model K). All other parameters for these two models are the same as indicated in
Table~\ref{tab:tableA}. As seen in the top panels of these figures, the CR distributions vary considerably for these values of $\cutoff$. The photon emission (bottom panels), of the trapped (dashed curves) and escaping CRs (solid curves) varies less strongly due to the fact that the photon emission is naturally spread out partially masking the shape of the underlying proton spectrum. One clear feature that remains is the low-energy kinematic cutoff at a few hundred MeV.
Of course, Figs.~\ref{fig:fpCutHalf} and \ref{fig:fpCutTwo} were calculated for a particular set of parameters and the relative importance of photon emission from escaping CRs versus trapped CRs will depend strongly on these parameters.

\section{Discussion and Conclusions}
As part of a comprehensive model of an evolving SNR undergoing
efficient
CR production, we have presented a \mc\ technique that describes the
diffusion of CRs that escape from the \FS\ of the remnant and
propagate
into a dense, external shell.
While a number of calculations of escaping CRs and their \gamray\
production have been performed \citep[see, for example,][ and
references therein]{LKE2008,OhiraEtal2010,Drury2010}, there remain many unresolved issues for this important problem. Our \mc\ method makes different assumptions than analytic calculations based on solving a diffusion equation and in some ways is less restrictive, particularly if
energy losses are included during propagation.

\begin{figure}
\epsscale{0.9}
\plotone{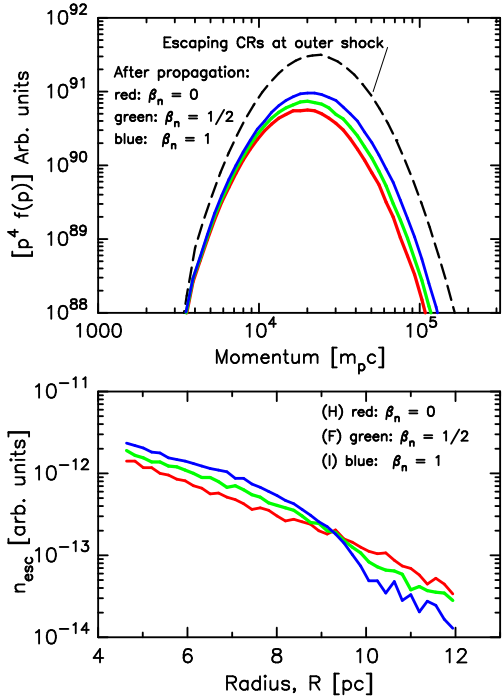} 
\caption{Escaping CR distributions, $\Fesc$, (top panel), and density
profiles (bottom panel) for Models H, F, and I, as listed
in Table~\ref{tab:tableA}. Note that the green curves (Model F) are
identical in Figs.~\ref{fig:vary_gyro} and \ref{fig:vary_density}. The
index $\nIndex$ is varied as shown and $\rgIndex=0.5$ in all cases.
\label{fig:vary_density}}
\end{figure}

The important features of our model include:
\newlistroman\listromanDE the energy content of the escaping CR
distribution is determined with the shock accelerated CRs
that remain trapped within the SNR using a 
planar, stationary, 
\NL\ model of efficient \DSA\ that neglects dilution
\citep[i.e.,][]{BGV2005};
\listromanDE the acceleration of CRs produces changes in the
hydrodynamics that modifies  the evolution of the SNR;
\listromanDE the shape of the trapped
CR distribution at the highest energies,
which is uncertain due to a lack of a well developed theory of
turbulence generation for anisotropic particles, is parameterized
consistently with the shape of the escaping CR distribution;
\listromanDE the broad-band continuum
photon emission from escaping and trapped
CRs is determined with a single set of environmental and model
parameters; and
\listromanDE although not emphasized or shown in the plots here, the thermal X-ray emission is included consistently with the broad-band continuum emission \citep[e.g.,][ and references therein]{EPSR2010}.

The examples we show indicate the complexity and importance of including escaping CRs in a consistent fashion with CRs that remain trapped within the SNR.
The shape of the GeV-TeV emission,  particularly the low-energy kinematic cutoff, is important as one of the main ways of determining whether this emission is \pion\ or \IC. If other features are discernable, they may provide clues to the importance of the escaping CRs and external density enhancements. We note that all of the spectra shown here are integrated over the region between the \CD\ and the \FS\ and are not line-of-sight projections. It should be clear from
Fig.~\ref{fig:ExtVary} that line-of-sight projections might show additional strong effects as escaping CRs interact with nearby dense material. Line-of-sight projections will be included in future work.

An important parameter that we haven't varied here is the efficiency
of DSA. In all of our examples we set $\EffDSA=50$\%, i.e., 50\% of
the forward shock ram kinetic energy flux goes into CRs (trapped and
escaping) at any instant. In fitting an actual SNR, $\EffDSA$ is a
parameter that may or may not be constrained by the observations. A
considerable amount of work has led to the conclusion that
$\EffDSA \sim 50$\% is a likely figure for young SNRs but this efficiency
will definitely vary between remnants, may vary during the remnant
lifetime, and may even vary at different locations in a single SNR
\citep[see, for example,][]{VBK2003}.
We note that \mc\ shock simulations  that include MFA and have parameters typical of young SNRs
\citep[e.g.,][]{VEB2006,VBE2008}, show total acceleration
efficiencies $\EffDSA \ge 50$\% with a sizable fraction of total shock ram kinetic energy ($\ge 30$\%) placed in escaping CRs.

Since we set $\EffDSA=50$\% for all of our examples, and
the other SNR parameters that determine what fraction of explosion
energy ends up in CRs are kept constant,
the third panel
in Fig.~\ref{fig:pmax} gives the results for all of our models. 
After 1000\,yr, $\sim 30$\% of the supernova explosion
energy has gone into all CRs with $\sim 10$\% going into escaping
CRs. At 10,000\,yr, $\sim 50$\% has gone into all CRs with $\sim
20$\% going into escaping CRs.

In this initial presentation of our \mc\ technique, we have exploded the supernova in a uniform CSM with an external, spherically symmetric shell of dense material. This simple scenario shows how important escaping CRs can be for modeling
non-thermal emission of young
SNRs. It is  not meant to match any particular object.
The \MC\ propagation part of the CR-hydro model can be easily generalized to include 
asymmetric external mass distributions, such as those expected when remnants interact with a dense molecular cloud (e.g., \SNRJ).
Future work will also model \gamrays\ produced when escaping CRs interact with the
complex structure of a dense surrounding shell as
expected from a 
progenitor stellar wind.

\acknowledgments We thank P. Slane for helpful discussions and A.
Vladimirov for calculating the escaping CR distribution from the
\citet{ZP2008} model. We also thank the referee, E. Berezhko, for helpful comments.
D.C.E. acknowledges support from NASA grants
ATP02-0042-0006, NNH04Zss001N-LTSA, and 06-ATP06-21. 
A.M.B. was
supported in part by the Russian government grant 11.G34.31.0001 through the Laboratory of Astrophysics with Extreme
Energy Release at St. Petersburg State Politechnical University,
RBRF
grants 09-02-12080, 11-02-00429, and by the RAS Presidium Program.
He performed some of the simulations at the Joint Supercomputing
Centre (JSCC RAS) and the Supercomputing Centre at Ioffe Institute,
St. Petersburg. The authors are grateful to the KITP in Santa
Barbara where part of this work was done when the authors were
participating in a KITP program.


\clearpage

\begin{table}
\begin{center}
\caption{Parameters for DSA and \CSM\ diffusion.} \label{tab:tableA}
\vskip6pt
\begin{tabular}{crrrrrrrrrrrr}
\tableline
\tableline
\\
Model
&$\Kep$
&$\cutoff$
&$\DCSMz$
&$\Lz$
&$\rgIndex$
&$\nIndex$
&$\Nuni$
&$\Nshell$
&$\Mshell$
&$\Rshell$
\\
&
&
& [cm$^2$s$^{-1}$]
& [pc]
&
&
& [\pcc]
& [\pcc]
& [$\Msun$]
& [pc]
\\
\\
\tableline
\\
A
& $1\xx{-2}$
& 1
& $1\xx{27}$
& $3.3\xx{-2}$
& 0.5
& 0.5
& 0.1
& 100
& $10^4$
& 10
\\
B
& $1\xx{-2}$
& 1
& $1\xx{27}$
& $3.3\xx{-2}$
& 0.5
& 0.5
& 0.1
& ---
& ---
& ---
\\
C
& $1\xx{-2}$
& 1
& $1\xx{26}$
& $3.3\xx{-3}$
& 0.5
& 0.5
& 0.1
& 100
& $10^4$
& 10
\\
D
& $1\xx{-2}$
& 1
& $1\xx{28}$
& $3.3\xx{-1}$
& 0.5
& 0.5
& 0.1
& 100
& $10^4$
& 10
\\
E
& $1\xx{-2}$
& 1
& $1\xx{27}$
& $3.3\xx{-2}$
& 1/3
& 0.5
& 1
& 10
& $10^3$
& 7
\\
F
& $1\xx{-2}$
& 1
& $1\xx{27}$
& $3.3\xx{-2}$
& 0.5
& 0.5
& 1
& 10
& $10^3$
& 7
\\
G
& $1\xx{-2}$
& 1
& $1\xx{27}$
& $3.3\xx{-2}$
& 1
& 0.5
& 1
& 10
& $10^3$
& 7
\\
H
& $1\xx{-2}$
& 1
& $1\xx{27}$
& $3.3\xx{-2}$
& 0.5
& 0
& 1
& 10
& $10^3$
& 7
\\
I
& $1\xx{-2}$
& 1
& $1\xx{27}$
& $3.3\xx{-2}$
& 0.5
& 1
& 1
& 10
& $10^3$
& 7
\\
J
& $1\xx{-4}$
& 0.5
& $1\xx{27}$
& $3.3\xx{-2}$
& 0.5
& 0.5
& 1
& 10
& $10^3$
& 7
\\
K
& $1\xx{-4}$
& 2
& $1\xx{27}$
& $3.3\xx{-2}$
& 0.5
& 0.5
& 1
& 10
& $10^3$
& 7
\\
\tableline
\tableline
\end{tabular}
\end{center}
\end{table}

\end{document}